\documentclass{article}
\usepackage{spconf,amsmath,graphicx}
\usepackage{booktabs}
\usepackage{makecell}

\usepackage{url}

\title{THE IDLAB VOXSRC-20 SUBMISSION: LARGE MARGIN FINE-TUNING AND QUALITY-AWARE SCORE CALIBRATION IN DNN BASED SPEAKER VERIFICATION}

\name{Jenthe Thienpondt, Brecht Desplanques, Kris Demuynck}
\address{IDLab, Department of Electronics and Information Systems, Ghent University - imec, Belgium}
\begin{document}
\ninept

%
\maketitle
\begin{abstract}
In this paper we propose and analyse a large margin fine-tuning strategy and a quality-aware score calibration in text-independent speaker verification. Large margin fine-tuning is a secondary training stage for DNN based speaker verification systems trained with margin-based loss functions. It enables the network to create more robust speaker embeddings by enabling the use of longer training utterances in combination with a more aggressive margin penalty. Score calibration is a common practice in speaker verification systems to map output scores to well-calibrated log-likelihood-ratios, which can be converted to interpretable probabilities. By including quality features in the calibration system, the decision thresholds of the evaluation metrics become quality-dependent and more consistent across varying trial conditions. Applying both enhancements on the ECAPA-TDNN architecture leads to state-of-the-art results on all publicly available VoxCeleb1 test sets and contributed to our winning submissions in the supervised verification tracks of the VoxCeleb Speaker Recognition Challenge 2020.

\end{abstract}
\section{Introduction}
\label{sec:intro}


Speaker verification solves the task whether two utterances are spoken by the same person. A recent shift towards neural network based speaker verification systems resulted in significantly better performance compared to the more traditional i-vector based systems \cite{x_vectors, x_vector_wide}. Current speaker verification systems consist of a Deep Neural Network (DNN), initially trained to classify utterances of a large number of training speakers. The most popular architectures are Time Delay Neural Networks (TDNN)~\cite{tdnn} and ResNet~\cite{resnet} based systems. A powerful improvement is the incorporation of an angular margin penalty to the standard softmax classification loss~\cite{arcface}. The statistics pooling layer that maps the variable length input to a fixed-length representation can be enhanced through a temporal attention mechanism~\cite{self_att, att_stat}. After training the network on the speaker identification task, fixed-length speaker characterizing embeddings can be constructed from the activations of the penultimate layer in the network. Subsequently, these embeddings can be used to score a speaker verification trial between new speakers. The most straightforward scoring method is the evaluation of the cosine distance between the enrollment and test embedding of the trials. An alternative Bayesian scoring technique is Probabilistic Linear Discriminant Analysis (PLDA) \cite{plda}. Often, this is followed by a score normalization step such as adaptive s-norm \cite{s_norm_2, s_norm_3}. Finally, logistic regression based score calibration can be used to map the output scores to reliable log-likelihood-ratios~\cite{bosaris}.


In this work, we increase the discriminative power of the neural network embedding extractor by introducing a large margin fine-tuning strategy. We show that using longer training segments allow the use of a larger margin penalty in the Additive Angular Margin (AAM) \cite{arcface} loss, which in its turn avoids the expected overfitting to the training speakers. This fine-tuning approach increases the inter-speaker distances between the more reliable speaker centers, while simultaneously ensuring compact speaker classes. We further enhance the speaker verification performance with a quality-aware calibration method. This logistic regression based calibration method is able to model various conditions of the trial utterances by including quality metrics. This results in more consistent speaker similarity scores across a wide range of conditions, and thus better performance given a single decision threshold in the evaluation metrics.


The paper is organized as follows: Section 2 describes our baseline speaker verification system. Section 3 and 4 will explain and motivate our proposed large margin fine-tuning and quality-aware calibration strategies respectively. Section 5 describes the experimental setup to analyse the proposed methods, while Section 6 explains and analyzes the results. Finally, Section 7 will give some concluding remarks.

\section{Baseline system}
\label{sec:baseline}

\begin{figure}[h]
\begin{minipage}[h]{1.0\linewidth}
  \centering
  \centerline{\includegraphics[scale=0.28]{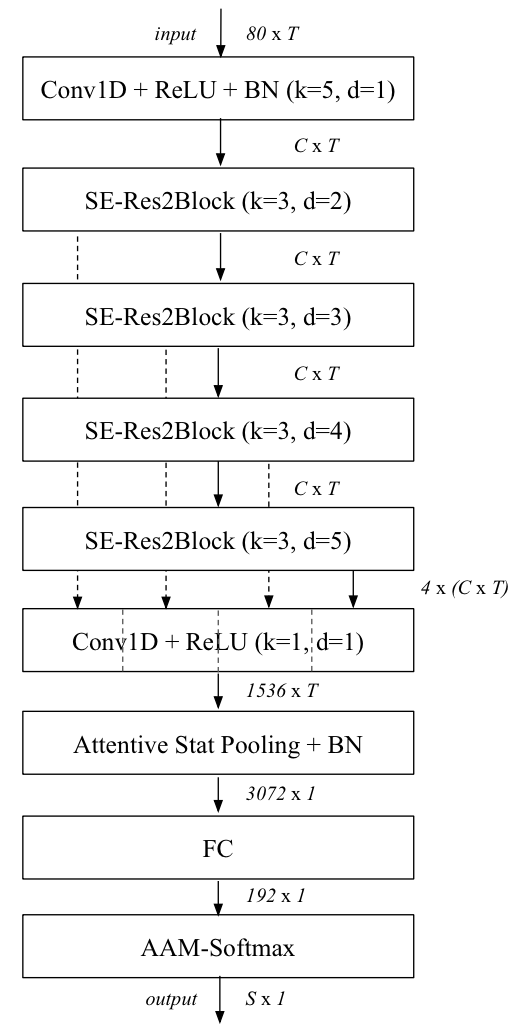}}
\end{minipage}

\caption{ECAPA-TDNN baseline system architecture. $T$ indicates the number of input frames, $C$ the amount of intermediate feature channels and $S$ the number of classification speakers. We denote \textit{k} and \textit{d} in the SE-Res2Block for the kernel size and dilation factor, respectively. See~\cite{ecapa_tdnn} for more details.}
\label{fig:res}
\end{figure}

Building further on our previously established work, we use the ECAPA-TDNN architecture~\cite{ecapa_tdnn} as our baseline speaker verification system. This TDNN based speaker embedding extractor improves the popular x-vector architecture~\cite{x_vectors} by incorporating a channel-and context-dependent attention system in the statistics pooling layer. It also introduces a 1-dimensional variant of Squeeze-Excitation (SE) blocks~\cite{se_block} to inject global information in frame-level layers of the network. The integration of Res2-blocks~\cite{res2net} improves performance while simultaneously reducing the total parameter count. Finally, multi-layer feature aggregation and feature summation allows the network to efficiently exploit knowledge learned in the preceding layers. The complete architecture is depicted in Figure~\ref{fig:res}, more details can be found in the ECAPA-TDNN publication~\cite{ecapa_tdnn}. We scale up the network compared to~\cite{ecapa_tdnn} by using 2048 feature channels and add a fourth SE-Res2Block with a dilation factor of 5 for optimized verification performance.


\section{Large margin fine-tuning}
\label{sec:fine-tuning}

One of the most successful loss functions in fine-grained classification and verification problems is AAM-softmax.
AAM-softmax is an extension of the standard softmax loss function by introducing an $L_2$-normalization step on the embeddings and applying an angular margin penalty during the estimation of the log-likelihood of the target class during training. This enforces the network to increase inter-speaker distances while ensuring intra-speaker compactness. The AAM-softmax loss is given by:
\begin{equation}
\label{aam_softmax}
L = -\frac{1}{n} \sum^{n}_{i=1} log \frac{e^{s(cos(\theta_{y_{i}} +m ))}}{e^{s(cos(\theta_{y_{i}} +m))} + \sum_{j=1, j \neq y_i}^N e^{s(cos(\theta_{j}))}}
\end{equation}
with $n$ indicating the batch size, $\theta_{y_i}$ the angle between the current embedding $\textbf{x}_{i}$ and the AAM-softmax class prototype $\textbf{y}_{i}$ with speaker identity $y_i$. The margin penalty is indicated with $m$. A scaling factor $s$ is applied to increase the range of the output log-likelihoods.

Higher values of $m$ result in more compact classes with larger inter-class distances, which should allow the network to better characterize differences between speakers. However, initial training of the network with a high margin penalty is difficult and often leads to poor results. Therefore it is common to train the network with a lower and probably sub-optimal margin. We propose a large margin fine-tuning strategy which further refines a network that was trained to convergence with a low margin value of $m$. Several changes on the level of the data preprocessing, data sampling and learning rate scheduling are proposed to stabilize and to optimize the fine-tuning stage at high margin settings.


\subsection{Fine-tuning configuration}
\label{ssec:fine-tuning_config}

During fine-tuning, we increase the duration of the training utterances. Most neural network based speaker verification systems are trained with short random temporal crops of 2 to 3 seconds to prevent overfitting. In this training configuration high margin penalties are too challenging and longer training sequences alleviate this issue by offering more speaker-specific information to the system. Moreover, this can potentially correct the duration mismatch between training and test conditions~\cite{magneto}. An effective method to decrease GPU memory requirements and to prevent overfitting when training with longer length utterances is to freeze the pre-pooling layers of the model~\cite{garcia_fine_tune}. However, we argue this can prevent these layers from sufficiently adapting to the increased duration condition, especially when such layers share global statistics through the SE-blocks in the ECAPA-TDNN architecture. Thus, all weights in the network remain trainable during the fine-tuning stage. To further prevent overfitting, we switch to a Cyclical Learning Rate (CLR) schedule~\cite{clr} with a lower maximum learning rate and shorter cycle-length.

We also enable the Hard Prototype Mining (HPM) sampling algorithm~\cite{sdsvc_paper} to create mini-batches with utterances from speakers which confuse the model the most. This speaker confusion is modelled through a speaker similarity matrix constructed by calculating the cosine distance between all AAM speaker prototype pairs. The sampling algorithm construct mini-batches by iterating randomly over all $N$ training speakers. Each iteration determines $S$ speakers, irrespective of their similarity, for which $U$ random utterances are sampled from each of their $I$ most similar speakers, including the selected speaker. This implies that $S \times U \times I$ should be equal to the batch size $n$. When we have iterated over all training speakers, the similarity matrix is updated and the batch generating process is repeated.

\section{Quality-Aware Score Calibration}
\label{sec:quality-aware_score_calibration}

Score calibration is a post-processing step in speaker verification systems to map similarity output scores to log-likelihood-ratios that can be converted to interpretable probabilities~\cite{bosaris}.
Well-calibrated scores allow a theoretical estimation of the optimal evaluation metric decision threshold on a wide range of possible decision error costs and prior probabilities of target and non-target verification trials. It also allows for easy score fusion by producing a weighted average across the calibrated system scores~\cite{bosaris}.

Research indicates that including quality measurements in the calibration stage can make the output scores more robust towards score-shifting caused by variability in recording quality and duration conditions \cite{bosaris, quality_cal_2013, quality_cal_2016}. However, most of this work is established with i-vector based speaker verification systems. We argue neural network based systems can benefit from quality measurements in the calibration step as well. Especially in the case of varying duration conditions, since most of these systems are trained with fixed-length audio crops for computational efficiency.

Calibration can be based on logistic regression which learns a weight $w_{s}$ and bias $b$ from a set of calibration trials to scale and shift the original output score to obtain a log-likelihood-ratio $l(s)$. However, this corresponds with a monotonic mapping from the score $s$ to $l(s)$. As a result, single system calibration  will not improve speaker verification performance metrics based on a fixed decision threshold, such as EER and MinDCF. The proposed quality-aware calibration mapping from input score $s$ to log-likelihood-ratio $l(s)$ is represented by:
\begin{equation}
\label{standard_cal}
l(s) = w_{s}s + \textbf{w}_{q}^{T} \textbf{q} + b
\end{equation}
By introducing additional quality measurements $\textbf{q}$ with learnable weights $\textbf{w}_{q}$, the calibration system becomes able to improve verification performance as the decision threshold implicitly becomes condition dependent.
The next subsections will describe and motivate various quality measurements included in our analysis.

\subsection{Duration-based quality measures}
\label{ssec:duration}

The most straightforward quality measurement is the duration of the utterance, which we will define as the number of input frames. The longer the audio input, the more confident the speaker verification system should be about its decision. However, the fraction of relevant speech frames could vary between utterances. Therefore, we also consider the amount of speech frames detected by the Voice Activity Detection (VAD) preprocessing module of SPRAAK~\cite{spraak} as a quality measure. Optionally, duration-based measurements can be clipped to a maximum length to reduce the impact on unexpectedly long recordings.

\subsection{Embedding-based quality measures}
\label{ssec:magnitude}

The magnitude of an embedding generated by a speaker embedding extractor trained with a softmax-based loss function could contain quality information about the original utterance \cite{ring_loss}. Small magnitudes could potentially indicate lower quality embeddings. Additionally, small changes on the embeddings close to the origin can have a big impact on the angles with the speaker prototypes.

\subsection{Imposter-based quality measures}
\label{ssec:imposter}

Prior to calibration, score normalization is often used to enhance speaker verification performance. The most common technique is s-normalization, which calculates a mean $\mu$ and standard deviation $\sigma$ of the distances between a test embedding and an imposter cohort. Adaptive s-normalization improves performance further by restricting the cohort to the most similar imposter speakers of the considered embedding.

The expected mean imposter score $\mu$ can be used as a quality metric, as the test utterance noise characteristics will have an impact on this similarity score. Moreover, a high $\mu$ indicates the system is wrongly confusing the test speaker with some of the training speakers and is probably not modelling the unseen speaker properly. However, experiments indicate that the magnitudes of the test embeddings play a crucial role. The average inner product between the test embedding and the speaker embeddings of its corresponding adaptive s-norm imposter cohort outperforms the average cosine similarity as an imposter-based quality metric~\cite{voxsrc_2020_technical_report}. Further research is required to analyze this interaction between embedding magnitudes and embedding cosine similarity.

\subsection{Symmetric Quality Measure Functions (QMFs)}
\label{ssec:side}

Since we want the role of the enrollment utterance and test utterance to be interchangeable during the verification trial, we enforce symmetric Quality Measure Functions (QMFs). The most straightforward way of combining the quality measurements of both utterances is by taking the arithmetic mean. However, this could result in the loss of valuable quality information as the output similarity score is potentially affected the most by the lowest-quality side of the trial. A simple solution would be to only consider the minimum quality measurement value along both sides of the trial. By also adding the maximum of the measurements as a separate feature we give the system the potential to model the quality difference between two utterances.

\begin{table*}
  \centering
  \begin{tabular}{clcccccccc}
    \toprule
    \multicolumn{1}{c}{} &
    \multicolumn{1}{l}{\textbf{System Configuration}} &
    \multicolumn{2}{c}{\textbf{VoxCeleb1}} &
    \multicolumn{2}{c}{\textbf{VoxCeleb1-E}} & 
    \multicolumn{2}{c}{\textbf{VoxCeleb1-H}} &
    \multicolumn{2}{c}{\textbf{VoxSRC-20 Val}} \\
    
    \cmidrule(lr){3-4} \cmidrule(lr){5-6} \cmidrule(lr){7-8} \cmidrule(lr){9-10}
    \multicolumn{2}{c}{\textbf{}} & 
    \multicolumn{1}{c}{\textbf{EER(\%)}} & \multicolumn{1}{c}{\textbf{MinDCF}} &
    \multicolumn{1}{c}{\textbf{EER(\%)}} & \multicolumn{1}{c}{\textbf{MinDCF}} &
    \multicolumn{1}{c}{\textbf{EER(\%)}} & \multicolumn{1}{c}{\textbf{MinDCF}} &
    \multicolumn{1}{c}{\textbf{EER(\%)}} & \multicolumn{1}{c}{\textbf{MinDCF}}\\
    
    \midrule
     & ECAPA-TDNN (C=2048) & 0.86 & 0.0960 & 1.08 & 0.1223 & 2.01 & 0.2004 & 3.25 & 0.2646\\
     & ECAPA-TDNN (fine-tuned) & 0.68 & 0.0753 & 0.91 & 0.1006 & 1.72 & 0.1695 & 2.89 & 0.2274\\
    \midrule
    \midrule
    A.1 & + duration QMF & 0.64 & 0.0764 & 0.88 & 0.0970 & 1.65 & 0.1638 & 2.68 & 0.2226 \\
    A.2 & + speech duration QMF & 0.63 & 0.0760 & 0.88 & 0.0970 & 1.64 & \textbf{0.1631} & 2.67 & 0.2218 \\
    A.3 & + magnitude QMF & 0.66 & 0.0765 & 0.90 & 0.1009 & 1.67 & 0.1694 & 2.87 & 0.2268 \\
    A.4 & + imposter mean QMF & 0.64 & \textbf{0.0700} & 0.89 & 0.1001 & 1.65 & 0.1724 & 2.81 & 0.2257 \\
    \midrule
    A.5 & + \makecell{speech duration QMF \& \\ imposter mean QMF}  & \textbf{0.56} & 0.0743 & \textbf{0.84} & \textbf{0.0969} & \textbf{1.57} & 0.1644 & \textbf{2.59} & \textbf{0.2185}\\
    
    \bottomrule
  \end{tabular}
    \caption{Performance impact of large margin fine-tuning and quality-aware score calibration on the ECAPA-TDNN system.}
  \label{tab:exp_results}
\end{table*}

\section{Experimental setup}
\label{sec:exp_setup}

We train the ECAPA-TDNN baseline model on the development set of the popular VoxCeleb2 dataset~\cite{vox2}. This training set contains over one million utterances across 5994 different speakers. We also create 6 additional augmented copies using the MUSAN~\cite{musan} corpus (babble, noise), the RIR~\cite{rirs} (reverb) dataset and the SoX (tempo up, tempo down) and FFmpeg (compression) libraries. 

The system is trained on random crops of 2~s to prevent overfitting. The input features are 80-dimensional MFCCs with a window size of 25~ms and a frame shift of 10~ms. To further improve robustness of the model, we apply SpecAugment \cite{specaugment} to the log mel-spectrograms which randomly masks 0 to 5 frames in the time-domain and 0 to 8 frequency bands. Subsequently, the MFCCs of the cropped segment are cepstral mean normalized. 

The initial margin penalty of the AAM-softmax layer is set to 0.2. We also apply a weight decay of 2e-5 on the weights in the network, except for the AAM-softmax layer, which uses a slightly higher value of 2e-4. The system is trained using the Adam optimizer \cite{adam} with a Cyclical Learning Rate (CLR) using the \textit{triangular2} policy as described in~\cite{clr}. The maximum and minimum learning rates are set at 1e-3 and 1e-8 respectively. We use a cycle length of 130k iterations with a batch size of 128. The model is trained for three full cycles.

We use adaptive s-normalization for all experiments in this paper. The imposter cohort consists of the average of the length-normalized utterance-based embeddings of each training speaker. The imposter cohort size is set to 100.

\subsection{Large margin fine-tuning setup}
\label{ssec:verification}

We apply our proposed large margin fine-tuning strategy on the converged baseline model. The margin of the AAM-softmax layer is increased to 0.5. SpecAugment is disabled and the length of the random crop is increased to 6~s, we noticed no further performance improvements by choosing a longer duration. The CLR cycle length is decreased to 60k, with the maximum learning rate lowered to \mbox{1e-5}. These shorter and less exploratory learning rate cycles should prevent the model from diverging too much from its initial starting position during fine-tuning. No layers in the network are frozen. Finally, the sampling strategy is changed to HPM as described in Section~\ref{ssec:fine-tuning_config} with parameter values $S = 16$, $I = 8$ and $U = 1$.
 
\subsection{Quality-aware calibration}
\label{ssec:exp_calibration}

To train our calibration system we create a set of trials from the VoxCeleb2 training dataset. We want our quality-aware calibration system to be robust against varying levels of duration in the trials. Subsequently, we define three types of trials: \textit{short-short}, \textit{short-long} and \textit{long-long} with \textit{short} indicating an utterance between 2 and 6 seconds and \textit{long} ranging from 6~s to the maximum length utterance in the VoxCeleb2 dataset. We include 10k trials of each type in our calibration trial set, resulting in a total of 30k trials. The amount of target and non-target trials is balanced.

\subsection{Evaluation protocol}
\label{ssec:eval_prot}

The proposed methods are evaluated on the public VoxCeleb1~\cite{vox1} test sets, which all have a similar duration distribution as the VoxCeleb2 training set. We also include system performance on the VoxSRC-20~\cite{voxsrc_2020_paper} validation set, which contains out-of-domain data. We report the EER and MinDCF metric using a $P_{target}$ value of $10^{-2}$ with $C_{FA}$ and $C_{Miss}$ set to 1. For the VoxSRC-20~\cite{voxsrc_2020_paper} test set results reported in Table~\ref{tab:ablation_voxsrc}, the MinDCF is evaluated as defined in the challenge with a $P_{target}$ value of $0.05$. Only the discriminatory ability of the system is evaluated by these metrics. We do not evaluate the actual calibration quality by e.g.\ ActDCF or $C_{llr}$~\cite{bosaris}.

\section{Results}
\label{sec:results}

In Table~\ref{tab:exp_results}, we show the performance impact of the large margin fine-tuning and quality-aware calibration on all VoxCeleb1 data sets. The ECAPA-TDNN baseline achieves strong results on all sets. Large-margin fine-tuning is beneficial, and results in an average relative improvement of 15\% in EER and 17\% in MinDCF. Experiments \textit{A.1-5} show that quality-aware calibration with the minimum and maximum QMF further improves these results. Evaluations \textit{A.1} and \textit{A.2} show that calibration with a duration-based QMF is very effective. The speech duration metric only delivers marginal gains compared to the basic total duration metric, indicating that the speech fraction of each utterance in VoxCeleb is rather consistent. Adding speech duration information during score calibration leads to an average improvement of 6\% in EER and 2\% in MinDCF across all datasets. The embedding magnitude is shown to be the weakest quality metric in experiment \textit{A.3}. However, it still improves performance which strengthens our belief to use the inner product as a similarity metric during the calculation of the imposter mean QMF in experiment \textit{A.4}. As shown in experiment \textit{A.5}, the speech duration and the imposter mean QMFs are clearly complementary and improve the performance by 11\% in EER and 3\% in MinDCF on average on all datasets compared to the fine-tuned baseline. We found no additional benefits by fusing more quality metrics or using QMFs with a logarithmic function.

\begin{table}[h]
  \centering
  \begin{tabular}{clcc}
    \toprule
     & \textbf{Systems} & \multicolumn{1}{c}{\textbf{EER(\%)}} & \multicolumn{1}{c}{\textbf{MinDCF}} \\
    \midrule
     &  No Fine-Tuning & 3.25 & 0.2646 \\
     &  Large Margin Fine-Tuning & \textbf{2.89} & \textbf{0.2274} \\
    \midrule
    \midrule
    B.1 &  No Margin Increase & 3.36 & 0.2672 \\
    B.2 &  No Duration Increase & 3.58 & 0.2884 \\
    B.3 &  No CLR Decrease & 4.87 & 0.3689 \\
    B.4 &  No Hard Sampling & 2.95 & 0.2345 \\
    B.5 &  Frozen Pre-Pooling Layers & 3.12 & 0.2399 \\
    \bottomrule
  \end{tabular}
    \caption{Ablation study of large margin fine-tuning on the \mbox{VoxSRC-20} validation set.}
    \label{tab:ablation_ft}
\end{table}

A detailed ablation study on our proposed large margin fine-tuning protocol is given in Table~\ref{tab:ablation_ft}. We observe significant relative performance improvements over the baseline system of 11\% in EER and 14\% in MinDCF on the VoxSRC-20 validation set. Experiments \textit{B.1-3} indicate the importance of combining the large margin setting with the selection of longer training utterances and reduction of the maximum learning rate in the CLR. In~\textit{B.3} the training becomes unstable and the results significantly degrade compared to the baseline. Hard sampling during fine-tuning is beneficial as indicated in experiment \textit{B.4}. Experiment \textit{B.5} shows that by freezing the pre-pooling layers during fine-tuning, potential performance gains are lost.

Extra experiments on the imposter mean QMF indicate that it is crucial to combine a top imposter cohort selection with the inner product speaker similarity metric. Imposter mean QMF variants that use cosine similarity or score against all training speakers do not deliver performance gains. For a concise ablation study of the proposed mean imposter QMF we refer to our technical report submitted to the VoxSRC-20 workshop~\cite{voxsrc_2020_technical_report}.


\begin{table}[h]
  \centering
  \begin{tabular}{lcc}
    \toprule
     \textbf{Systems} & \multicolumn{1}{c}{\textbf{EER(\%)}} & \multicolumn{1}{c}{\textbf{MinDCF}} \\
    \midrule
     Fusion & 4.20 & 0.2052 \\
     Fusion + Fine-Tuning & 4.06 & 0.1890 \\
     Fusion + Fine-Tuning + QMFs & \textbf{3.73} & \textbf{0.1772}  \\
    \bottomrule
  \end{tabular}
    \caption{Evaluation of the proposed fine-tuning and quality-aware calibration (QMFs) on the VoxSRC-20 test set.}
    \label{tab:ablation_voxsrc}
\end{table}

For a final evaluation of our proposed methods, we provide a result overview in Table~\ref{tab:ablation_voxsrc} of our winning submission in the VoxSRC-20 closed speaker verification track. The fusion system is a score-level fusion between 6 ECAPA-TDNN and 4 Resnet34 variants, see~\cite{voxsrc_2020_technical_report} for more details. Large-margin fine-tuning of all models results in a relative improvement of 3\% in EER and 8\% in MinDCF. Using quality-aware score calibration of the fused score with the speech duration and imposter mean QMFs resulted in an additional 8\% EER and 6\% MinDCF relative improvement.


\section{Conclusion}
\label{sec:conclusion}

Large margin DNN fine-tuning can result in the generation of more speaker discriminative embeddings, provided that longer training utterances and a reduced learning rate are used. In addition, quality-aware score calibration that uses quality metrics of a verification trial is able to produce more robust log-likelihood-ratios. Applying both techniques on an ECAPA-TDNN model resulted in state-of-the-art performance on all VoxCeleb1 test sets. Our submission with system fusion in the VoxSRC-20 competition delivered a relative improvement of 11\% in EER and 14\% in MinDCF by applying both enhancements. This approach resulted in a first-place finish on both supervised speaker verification tracks.


\vfill\pagebreak

\bibliographystyle{IEEEbib}
\bibliography{refs}

\end{document}